\documentclass[%
 reprint,
 amsmath,amssymb,
 aps,
]{revtex4-2}

\usepackage{graphicx}% Include figure files
\usepackage{dcolumn}% Align table columns on decimal point
\usepackage{bm}% bold math
\begin{document}

%\preprint{}

\title{Electrons in helical magnetic field: a new class of topological metals}% Force line breaks with \\
%\thanks{A footnote to the article title} 

\author{Yu. B. Kudasov}

\affiliation{Sarov Physics and Technology Institute NRNU MEPhI, 6 Dukhov str., Sarov, 607186, Russia}
\affiliation{Russian Federal Nuclear Center - VNIIEF, 37 Mira pr., Sarov, 607188, Russia}%

\date{\today}

\begin{abstract}
Two theorems on electron states in helimagnets are proved. They reveal a Kramers-like degeneracy in helical magnetic field. Since a commensurate helical magnetic system is transitionally invariant with two multiple periods (ordinary translations and generalized ones with rotations), the band structure turns out to be topologically nontrivial. Together with the degeneracy, this gives an unusual spin structure of electron bands. A 2D model of nearly free electrons is proposed to describe conductive hexagonal palladium layers under an effective field of magnetically ordered CrO$_2$ spacers in PdCrO$_2$. The spin texture of the Fermi surface leads to abnormal conductivity.     
\end{abstract}

\maketitle

Metallic delafossites  PtCoO$_2$, PdCoO$_2$, PdCrO$_2$ are layered compounds with anomalous transport properties \cite{Mackenzie,Akaike,Takatsu,Hicks}. Their conductivity at room temperature approaches to that of the best elementary conductors like argentum, copper, and aluminum \cite{Mackenzie}. CoO$_2$ and CrO$_2$ are proved to be dielectric interlayers \cite{Lechermann}, and the conductivity in these substances is determined entirely by hexagonal palladium or platinum layers. This leads to extremely large values of electron mean free path, up to 20~m$\mu$ at low temperatures \cite{Hicks}. When electron momentum-relaxing scattering by
impurities or phonons is much weaker as compared to momentum-conserving scattering, a hydrodynamic regime of electron transport appears \cite{Scaffidi} as it was observed in PdCoO$_2$ \cite{Moll}.
The unusual behavior of the delafossites suggests a novel mechanism of electron transport in them \cite{Usui}.

A long-range magnetic order appears in PdCrO$_2$ below $T_c = 37.5$~K \cite{Mekata,Takatsu2,Billington}. Chromium ions form a 120$^0$ magnetic structure within a single layer. A complex interlayer arrangement leads to the magnetic system consisting of 18 sublattices \cite{Takatsu2}. The appearance of the magnetic order is accompanied by a resistivity drop \cite{Daou}. Thus, a main hypothesis, which is going to be proved in the present article, is that the helical magnetic order on certain conditions induces the high-conductivity state. It also should be mentioned that unconventional anomalous Hall effect \cite{Takatsu3} and nonreciprocal electronic transport \cite{Akaike} are observed in PdCrO$_2$ in the magnetically ordered state. Above $T_c$ a short-range magnetic order in chromium hexagonal layers persists up to about 500~K \cite{Daou,Billington}, and this is another surprising fact in view of the extremely high conductivity.    

The Fermi surface in the metallic delafossites was thoroughly investigated \cite{Hicks,Mackenzie}. In the paramagnetic state, it is quasi-two-dimensional with a single $\alpha$-orbit. The transition to the ordered state in PdCrO$_2$ leads to a reconstruction of the Fermi surface within the magnetic Brillouin zone and appearance of additional $\gamma$-orbits corresponding to pockets in the vicinity of K points \cite{Billington,Hicks2}.  

A motion of spin-1/2 particle in helical magnetic field was studied for a long time \cite{Dzyaloshinskii,Calvo,Calvo2,Fresard,Fraerman,Kishine}. In particular, an exact solution is known \cite{Calvo} and there are various approximate approaches \cite{Calvo2,Kudasov}. The electron transport in noncollinear magnetic structures demonstrates nonreciprocal due to breaking of spatial inversion symmetry \cite{Akaike,Fraerman}. A spin space group (SSG) theory is a useful instrument for investigation of helical magnetic systems \cite{Brinkman,Sandratskii}. A SSG operator $\{\mathbf{ \boldsymbol\alpha}|\mathbf{ \boldsymbol\beta}|\mathbf{t}\}$ comprises the spin rotation $\mathbf{ \boldsymbol\alpha}_s$ and space transformation $\{\mathbf{ \boldsymbol\beta}|\mathbf{t}\}$ combining rotation $\mathbf{ \boldsymbol\beta}$ and translation $\mathbf{t}$. In the framework of the theory, generalized translations are introduced ($\{\mathbf{ \boldsymbol\alpha}|0|\mathbf{t}\}$), and a generalized Bloch theorem was proved \cite{Brinkman,Sandratskii}.

Topological aspects of band structure of crystalline solids were intensively studied during last decades \cite{Bansil}. Main efforts of theoreticians and experimenters were concentrated on investigations of topological insulators and their edge states \cite{Tkachev,Hasan}. Recently, magnetic topological insulators and semimetals including helimagnets turned out to be the focus of attention \cite{Bernevig,Yao}. In the present article, a novel approach to topology of metallic helimagnets is proposed.

Let us consider a particle of spin-1/2 in a nonuniform magnetic field, which is invariant under translation with period $\mathbf{T}$, i.e. $\mathbf{h}(\mathbf{r}+\mathbf{T})=\mathbf{h}(\mathbf{r})$. The Hamiltonian of the system has the form
\begin{eqnarray}
	\hat{H}=\hat{H}_0 + \mathbf{h}(\mathbf{r}) \hat{\mathbf{ \boldsymbol\sigma}} \label{H}
\end{eqnarray}
where $\hat{\mathbf{\boldsymbol\sigma}}$ are the Pauli matrices. One can prove two theorems on eigenvalues of the Hamiltonian.

{\bf{Theorem 1.}} If $\hat{H}_0$ is invariant under the operation $\hat{\mathbf{T}}_{1/2} \hat{\mathbf{\theta}}$, where $\hat{\mathbf{T}}_{1/2}$ is the space translation along vector $\mathbf{T}/2$, $\hat{\mathbf{\theta}}$ is the time-reversal operator, and $\mathbf{h}(\mathbf{r}+\mathbf{T}/2)=-\mathbf{h}(\mathbf{r})$, then eigenvalues $\varepsilon _ {\mathbf{k}}$ of $\hat{H}$ are at least two-fold degenerate for all the wave vectors $\mathbf{k}$ except those, which satisfy $\exp\big(i \mathbf{k} \mathbf{T}\big)=-1$, and the following condition is fulfilled for all the eigenvalues: 
\begin{eqnarray}
	\varepsilon _ {\mathbf{k}, \langle \boldsymbol\sigma \rangle} = \varepsilon _ {-\mathbf{k},- \langle \boldsymbol\sigma \rangle} \label{T1}
\end{eqnarray}
where $ \langle \boldsymbol\sigma \rangle \equiv \langle \psi_ {\mathbf{k}} | \hat{ \boldsymbol\sigma} | \psi_ {\mathbf{k}} \rangle$ and $| \psi_ {\mathbf{k}} \rangle $ is the eigenstate of $\hat{H}$.

{\bf{Proof.}}  In the presence of magnetic field a symmetry operation containing $\hat{\mathbf{\theta}}$ should be a product of an operation, which
change sign of the magnetic field, and $\hat{\mathbf{\theta}}$ \cite{Wigner}. Therefore, $\hat{\mathbf{T}}_{1/2} \hat{\mathbf{\theta}}$ is a symmetry operation for $\hat{H}$. Let us consider two eigenstates: $| \psi_ {\mathbf{k}} \rangle$ and $\hat{\mathbf{T}}_{1/2} \hat{\mathbf{\theta}} | \psi_ {\mathbf{k}} \rangle$. According to rules for antiunitary operators \cite{Messiah} we obtain
\begin{eqnarray}
	\langle \psi_ {\mathbf{k}} | \big( \hat{\mathbf{T}}_{1/2} \hat{\mathbf{\theta}} | \psi_ {\mathbf{k}} \rangle \big) =  \overline{\big(\langle \psi_ {\mathbf{k}} |\hat{\mathbf{\theta}}^\dagger \big) \big( \hat{\mathbf{T}}_{1/2} \hat{\mathbf{\theta}}^2 | \psi_ {\mathbf{k}} \rangle \big)} \nonumber \\ =- \langle \psi_ {\mathbf{k}} | \hat{\mathbf{T}}^\dagger  \big( \hat{\mathbf{T}}_{1/2} \hat{\mathbf{\theta}} | \psi_ {\mathbf{k}} \rangle \big).  \label{H11}
\end{eqnarray} 
By means of Bloch's theorem $ \hat{\mathbf{T}}| \psi_ {\mathbf{k}} \rangle = \exp\big(i \mathbf{k} \mathbf{T}\big)| \psi_ {\mathbf{k}} \rangle$ it is reduced to
\begin{eqnarray}
	\langle \psi_ {\mathbf{k}} | \big( \hat{\mathbf{T}}_{1/2} \hat{\mathbf{\theta}} | \psi_ {\mathbf{k}} \rangle \big) = -\exp\big(i \mathbf{k} \mathbf{T}\big) \langle \psi_ {\mathbf{k}} | \big( \hat{\mathbf{T}}_{1/2} \hat{\mathbf{\theta}} | \psi_ {\mathbf{k}} \rangle \big).  \label{H12}
\end{eqnarray}
From here one can see that $| \psi_ {\mathbf{k}} \rangle$ and $\hat{\mathbf{T}}_{1/2} \hat{\mathbf{\theta}} | \psi_ {\mathbf{k}} \rangle$ are orthogonal and make up a pair of degenerate states for all $\mathbf{k}$ except those, which satisfy $\exp\big(i \mathbf{k} \mathbf{T}\big)=-1$. In the last case,
the state can be either degenerate or nondegenerate. Since operators $\hat{\mathbf{k}}$ and $\hat{\sigma}$ commute with $\hat{\mathbf{T}}_{1/2}$ and change sign under the time reversal, we obtain Eq.~(\ref{T1}) $\blacksquare$. 

The first term of Eq.(\ref{H}) can contain a scalar potential and kinetic energy contribution $\hat{\mathbf{p}}^2/(2m)$ where $\hat{\mathbf{p}}$  and $m$ are the momentum operator and particle mass. If the vector potential corresponding to $\mathbf{h}(\mathbf{r})$ has the same translational symmetry  under a proper calibration, i.e. $\mathbf{A}(\mathbf{r}+\mathbf{T/2})=-\mathbf{A}(\mathbf{r})$, 
the kinetic term can be extracted from $\hat{H}_0$  and the Hamiltonian 
\begin{eqnarray}
	\hat{H} = \hat{H}_0 + \big[ \hat{\mathbf{p}} + q_0 \mathbf{A}(\mathbf{r}) \hat{\mathrm{I}} \big]^2/(2m)  + \mathbf{h}(\mathbf{r}) \hat{\mathbf{ \boldsymbol\sigma}}  \label{H00}
\end{eqnarray}
also meets the theorem. Here $q_0$ is the charge of the particle and $\hat{\mathrm{I}}$ is the unit matrix. 

Helical systems with the SSG symmetry operator $\{\mathbf{ \boldsymbol\alpha}|0|\mathbf{t}\}$, where $\alpha=2\pi/n$ and $n$ is an even natural number, satisfy the hypothesis of the theorem if $\mathbf{h}(\mathbf{r})$ is perpendicular to the axis of spin rotation. This can be illustrated by an example of a tight-binding model for a four-sublattice helical structure (see Supplemental material). On the other hand, 
the theorem can not be applied if $n$ is the odd number, in particular, in the important case of the 120$^0$ magnetic ordering ($n=3$).

Let us introduce an operator $\hat{\mathbf{r}}_{\alpha}$ of spin rotation by angle $\alpha$ about $z$ axis. In the theorem and models discussed below, there is no spin-orbit coupling. Therefore, the spin rotation axis can be chosen arbitrary, i.e. it is independent of the spacial $z^\prime$ axis.

{\bf{Theorem 2.}} If $\hat{H}_0$ in Eq.(\ref{H}) is invariant under translation $\hat{\mathbf{t}}$, time reversion $\hat{\mathbf{\theta}}$, and arbitrary rotations of the spin system about $z$ axis, and if $\mathbf{h}(\mathbf{r})$ is invariant under $\hat{\mathbf{r}}_{\alpha} \hat{\mathbf{t}}$, where  $\alpha=2\pi/n$, $n$ is an odd number ($n>1$), and $\mathbf{h}(\mathbf{r})$ is perpendicular to the spin rotation axis, then the eigenvalues of $\hat{H}$ are at least two-fold degenerate for all the wave vectors $\mathbf{k}$ except  those, which satisfy $\exp\big(2 i \mathbf{k} \mathbf{T}\big)=1$, where $\mathbf{T}=n\mathbf{t}$, and
all the eigenvalues satisfy Eq.~(\ref{T1}) with $\langle \hat{\mathbf{\sigma}}_{x(y)} \rangle_ {\mathbf{k}} = 0$.

{\bf{Proof.}}  Let us introduce operators    
\begin{eqnarray}
	\hat{\mathbf{Y}} = \hat{\mathbf{t}} \hat{\mathbf{r}}_{\alpha}  \mbox{   and   }	\hat{\mathbf{X}} = \hat{\mathbf{t}} \hat{\mathbf{r}}_{\alpha-\pi} \hat{\mathbf{\theta}}. \label{XY}
\end{eqnarray}
$\hat{\mathbf{Y}}$ is a symmetry operator under the hypothesis of the theorem and a generator of an Abelian group \cite{Sandratskii}. Its irreducible representations coincide with those of the space translation group $\exp(i\mathbf{k}\mathbf{t})$,
 where $\mathbf{k}$ is the wave vector in the extended Brillouin zone \cite{Sandratskii}. The magnetic field is perpendicular to the spin rotation axis, then $\hat{\mathbf{r}}_{-\pi}\mathbf{h}(\mathbf{r}) = - \mathbf{h}(\mathbf{r})$ and $\hat{\mathbf{X}}$ is also a symmetry operator. It forms a group, which is isomorphic to that of $\hat{\mathbf{Y}}$. We can consider the following quantity:
\begin{eqnarray}
	\langle \psi_ {\mathbf{k}} | \big( \hat{\mathbf{X}}^n | \psi_ {\mathbf{k}} \rangle \big) = 	\langle \psi_ {\mathbf{k}} | \big( \hat{\mathbf{X}}^{-n} | \psi_ {\mathbf{k}} \rangle \big).  \label{H21}
\end{eqnarray}
Direct translations give
\begin{eqnarray}
	\langle \psi_ {\mathbf{k}} | \big( \hat{\mathbf{t}}^{-n} \hat{\mathbf{r}}_{\alpha-\pi}^{-n} \hat{\mathbf{\theta}}^{-n} | \psi_ {\mathbf{k}} \rangle \big) =  \langle \psi_ {\mathbf{k}} | \hat{\mathbf{t}}^{-2n} \big( \hat{\mathbf{t}}^{n} \hat{\mathbf{r}}_{\alpha-\pi}^{-n} \hat{\mathbf{\theta}}^{-n} | \psi_ {\mathbf{k}} \rangle \big) \nonumber \\
	=  - \langle \psi_ {\mathbf{k}} | \hat{\mathbf{t}}^{-2n} \hat{\mathbf{r}}_{\alpha-\pi}^{-2n} \big( \hat{\mathbf{t}}^{n} \hat{\mathbf{r}}_{\alpha-\pi}^{n} \hat{\mathbf{\theta}}^{n} | \psi_ {\mathbf{k}} \rangle \big) .  \label{H22}
\end{eqnarray}
Using $\hat{\mathbf{r}}_{\alpha-\pi}^{-2n} = - \hat{\mathrm{I}}$ and Bloch's theorem $ \hat{\mathbf{T}}^2| \psi_ {\mathbf{k}} \rangle = \exp\big(2 i \mathbf{k} \mathbf{T}\big)| \psi_ {\mathbf{k}} \rangle$, we obtain
\begin{eqnarray}
	\langle \psi_ {\mathbf{k}} | \big( \hat{\mathbf{X}}^n | \psi_ {\mathbf{k}} \rangle \big) = \exp\big( - 2 i \mathbf{k} \mathbf{T}\big)	\langle \psi_ {\mathbf{k}} | \big( \hat{\mathbf{X}}^n | \psi_ {\mathbf{k}} \rangle \big).  \label{H23}
\end{eqnarray}
That is, $| \psi_ {\mathbf{k}} \rangle$ and $\hat{\mathbf{X}}^n | \psi_ {\mathbf{k}} \rangle$ are orthogonal and the eigenstates are two-fold degenerate for all the wave vectors except those, which satisfy $\exp\big(2 i \mathbf{k} \mathbf{T}\big)=1$. The operators $\hat{\mathbf{k}}$ and $\hat{\sigma}_z$ commute with $\hat{\mathbf{t}}$ and $\hat{\mathbf{r}}$, as well as they change sign under the time reversal. Therefore, we obtain 
$\varepsilon_{\mathbf{k},\langle \sigma _z \rangle} = \varepsilon_{-\mathbf{k}, - \langle \sigma _z \rangle}$. The relations for the transverse spin components are proved in Supplement material  $\blacksquare$.

\begin{figure*}
	 \hfill \includegraphics[width=0.5\textwidth]{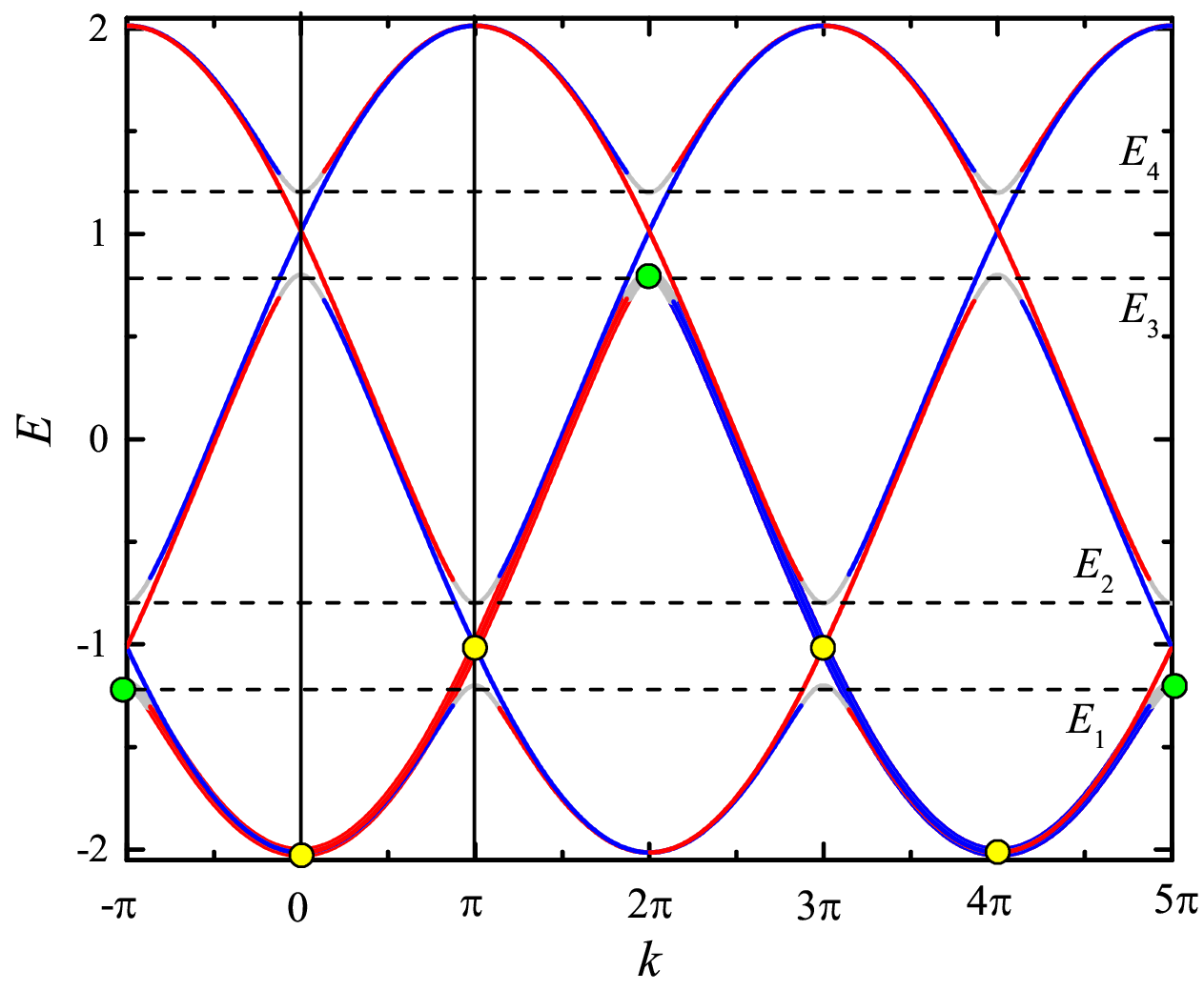} \hfill
	\includegraphics[width=0.25\textwidth]{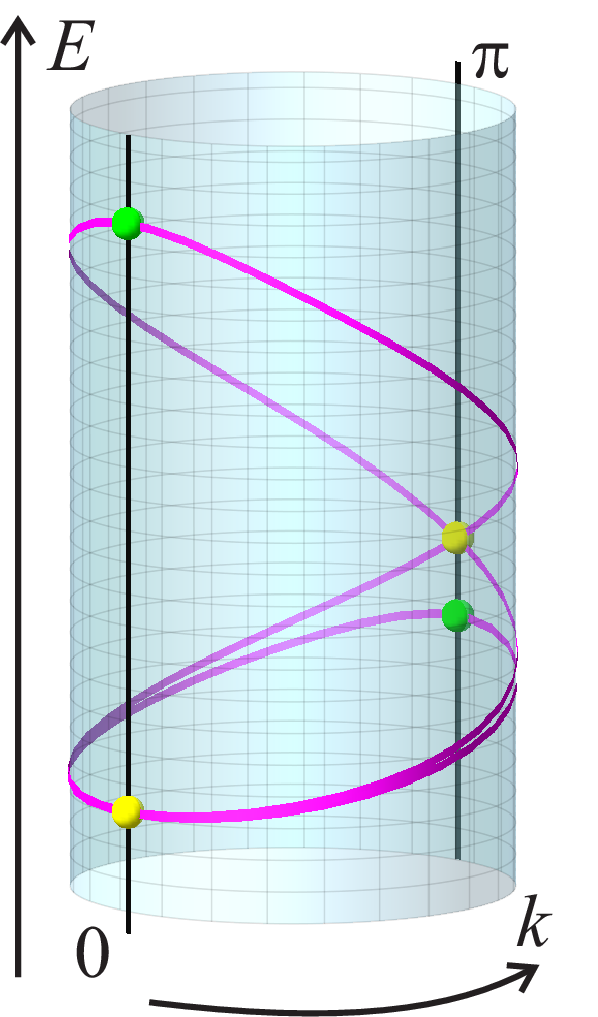}  \hfill
	\caption{\label{f1} The electron band structure of the 1D model Eq.~(\ref{H3}) and a schematic representation of the bottom dispersion curve on cylinder.  The average spin along the curves is indicated by color: red and blue if $|\langle \hat{\mathbf{\sigma}}_z\rangle|>1/2$, gray otherwise. The yellow and green circles denote the degenerate and non-degenerate points, correspondingly.}
\end{figure*}

If the vector potential has the same translational symmetry as $\mathbf{h}(\mathbf{r})$, i.e. $\hat{\mathbf{r}}_{\alpha} \hat{\mathbf{t}}\mathbf{A}(\mathbf{r})=\mathbf{A}(\mathbf{r})$, Hamiltonian of the form (\ref{H00}) also satisfies theorem 2.

Let us consider a tight-binding model of an atomic chain under an effective magnetic field corresponding to the 120$^0$ order as an example of 1D helical system satisfying to theorem 2:
\begin{eqnarray}
	\hat{H}_{3sl}=	-\sum_{i,\sigma}\left( \hat{a}^\dagger_{i,1,\sigma} \hat{a}_{i,2,\sigma} +\hat{a}^\dagger_{i,2,\sigma} \hat{a}_{i,3,\sigma} +\right. \nonumber\\ \left. \hat{a}^\dagger_{i,3,\sigma} \hat{a}_{i+1,1,\sigma} + h.c.\right) -
	\sum_{i,j,\sigma,\sigma^\prime}\left( \hat{a}^\dagger_{i,j,\sigma} \hat{\mathbf{h}}_{j} \hat{a}_{i,j,\sigma^\prime} \right).  \label{H3}
\end{eqnarray}
where $\hat{a}^\dagger_{i,j,\sigma} (\hat{a}_{i,j,\sigma} )$ is the electron creation (annihilation) operator in the $j$-th sublattice ($j=1,2,3$) and $i$-th cell with the spin projection on $z$ axis $\sigma=\pm 1/2$. The on-site magnetic field for the sublattices is defined as follows: $\hat{\mathbf{h}}_{1}=h_0 \hat{\sigma}_x$, $\hat{\mathbf{h}}_{2}=h_0 (-\hat{\sigma}_x+\sqrt{3}\hat{\sigma}_y)/2$, $\hat{\mathbf{h}}_{3}=h_0 (-\hat{\sigma}_x-\sqrt{3}\hat{\sigma}_y)/2$.

This model is exactly solvable and the electron dispersion is shown in Fig.~\ref{f1}. The magnetic Brillouin zone lies between $-\pi$ and $\pi$. According to theorem 2, the obtained eigenvalues obey Eq.~(\ref{T1}).
Special points, where the degeneracy is undefined ($k=0$ and $k=\pm\pi$), are shown by yellow and green circles.

As was mentioned above the helical system is subject to both the generalized Bloch theorem due to invariance under the generalized translation $\hat{\mathbf{r}}_{\alpha} \hat{\mathbf{t}}$ \cite{Brinkman,Sandratskii} and ordinary Bloch theorem \cite{Ashcroft} (translational invariance under  $\hat{\mathbf{T}}=\hat{\mathbf{t}}^n$). Therefore, the eigenstates and eigenvalues are periodic in the reciprocal space with two periods which are multiple of one another. 

We define dispersion curves so that the corresponding eigenstates  $| \psi_ {\mathbf{k},g} \rangle$ are continuous functions of $\mathbf{k}$ over the extended Brillouin zone defined by the generalized translations, and $g$ is the curve number. As an example, one of the curve is shown in Fig.~\ref{f1} by the thick line. According to the Bloch theorem, the eigenvalues should obey the conditions $\epsilon_{l,\mathbf{k}+\mathbf{K}}=\epsilon_{l,\mathbf{k}}$ where $\mathbf{K}$ is the primitive vector of the magnetic reciprocal lattice and $l$ is the eigenvalue number. One can see in Fig.~\ref{f1} that a dispersion curve is not necessarily periodic in the magnetic Brillouin zone. However, the whole band structure has to be periodic to fulfill the Bloch theorem. 

For instance, let us denote the lowest dispersion curve in the range from $-\pi$ to $0$ in Fig.~\ref{f1} as $| \psi_ {\mathbf{k},1} \rangle$. It corresponds to $l=1$ (eigenvalues indexed from the bottom to the top). However, while crossing $k=0$ and, then, $k=\pi$ the eigenvalue number of the curve is changed by $l=2$ and $l=3$. That is, there is no a single-valued correspondence between $g$ and $l$.

This leads to a non-trivial topology of dispersion curves. The curve marked by the thick line in the left panel of Fig.~\ref{f1} is schematically shown in a cylinder representation in the right panel. Paths on cylinder can be classified by means of the fundamental group of cylinder \cite{Malcev}. The winding number coincides with $n$ defined above and is a topological index. In the magnetic Brillouin zone, the band is reduced to $n$ branches ($n=3$ in Fig.~\ref{f1}).
If the Fermi level falls into a gap between $E_2$ and $E_1$ (or between $E_4$ and $E_3$) an electron transport becomes unusual. A backward scattering without spin flip is forbidden. There also exists a persistent spin current.

An additional uniform magnetic field directed along spin $z$ axis breaks the Kramers-like symmetry defined by theorem 2. On the other hand, the Hamiltonian remains still invariant under both the translations. That is why, the topological structure in the right panel of Fig.~\ref{f1} survives under the perturbation. A uniform magnetic field perpendicular to the $z$ axis breaks both the Kramers-like symmetry and invariance under the generalized translations. In this case, the topology becomes trivial due to hybridization at band crossing points.

\begin{figure*}
	\includegraphics[width=0.35\textwidth]{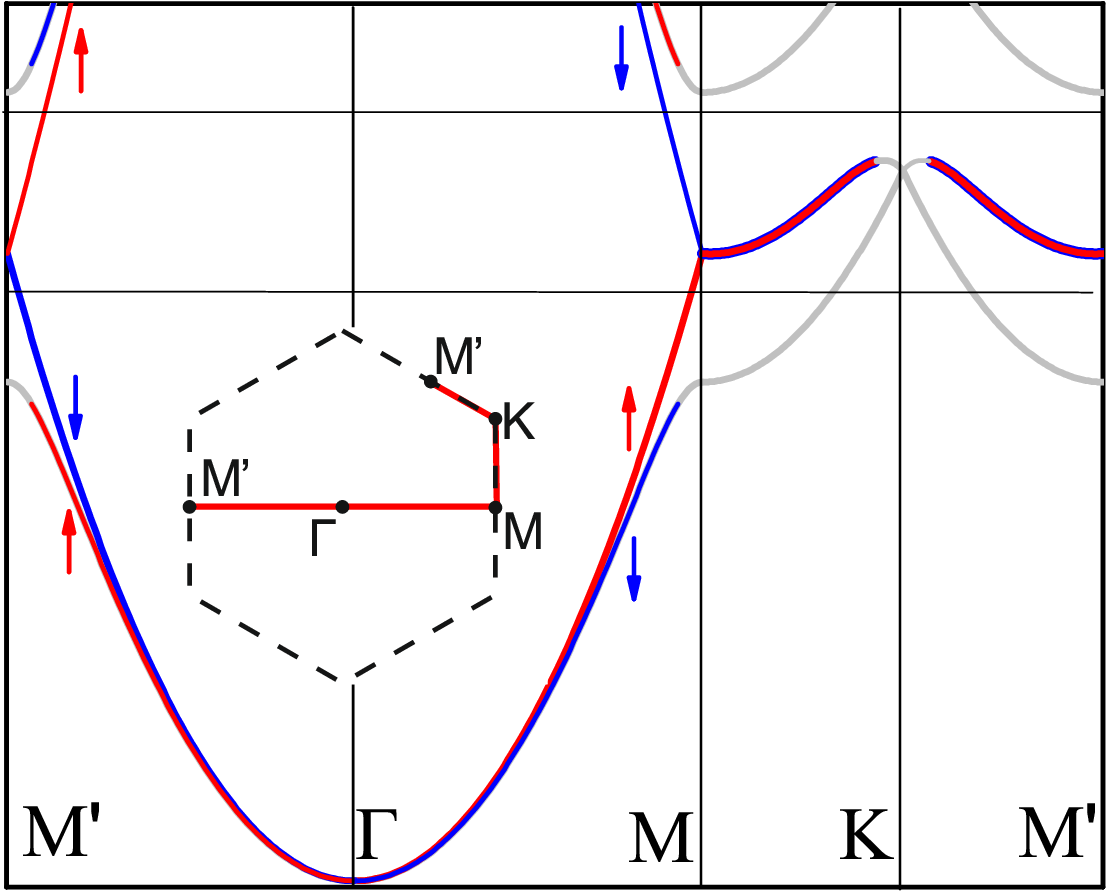} \hfill
	\includegraphics[width=0.28\textwidth]{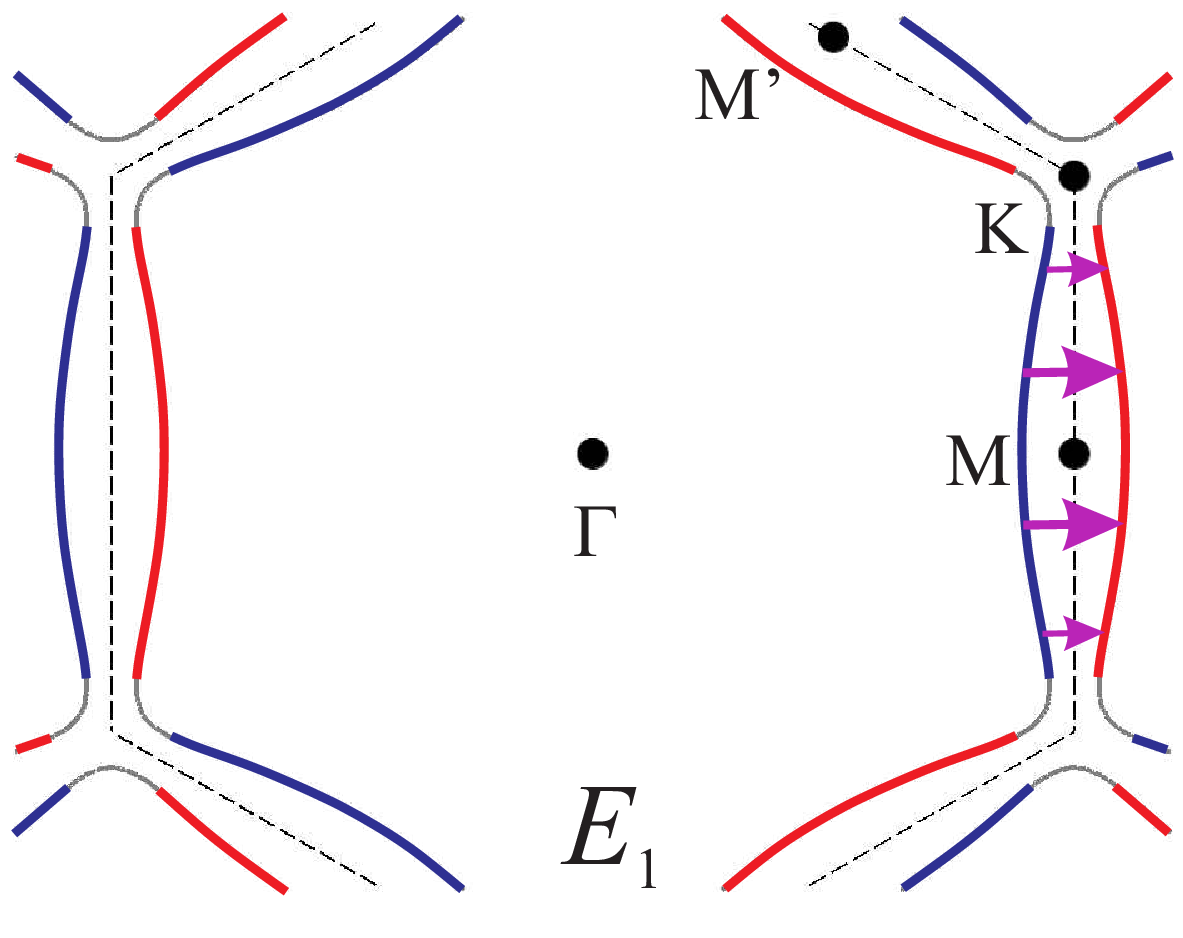} \hfill
	\includegraphics[width=0.28\textwidth]{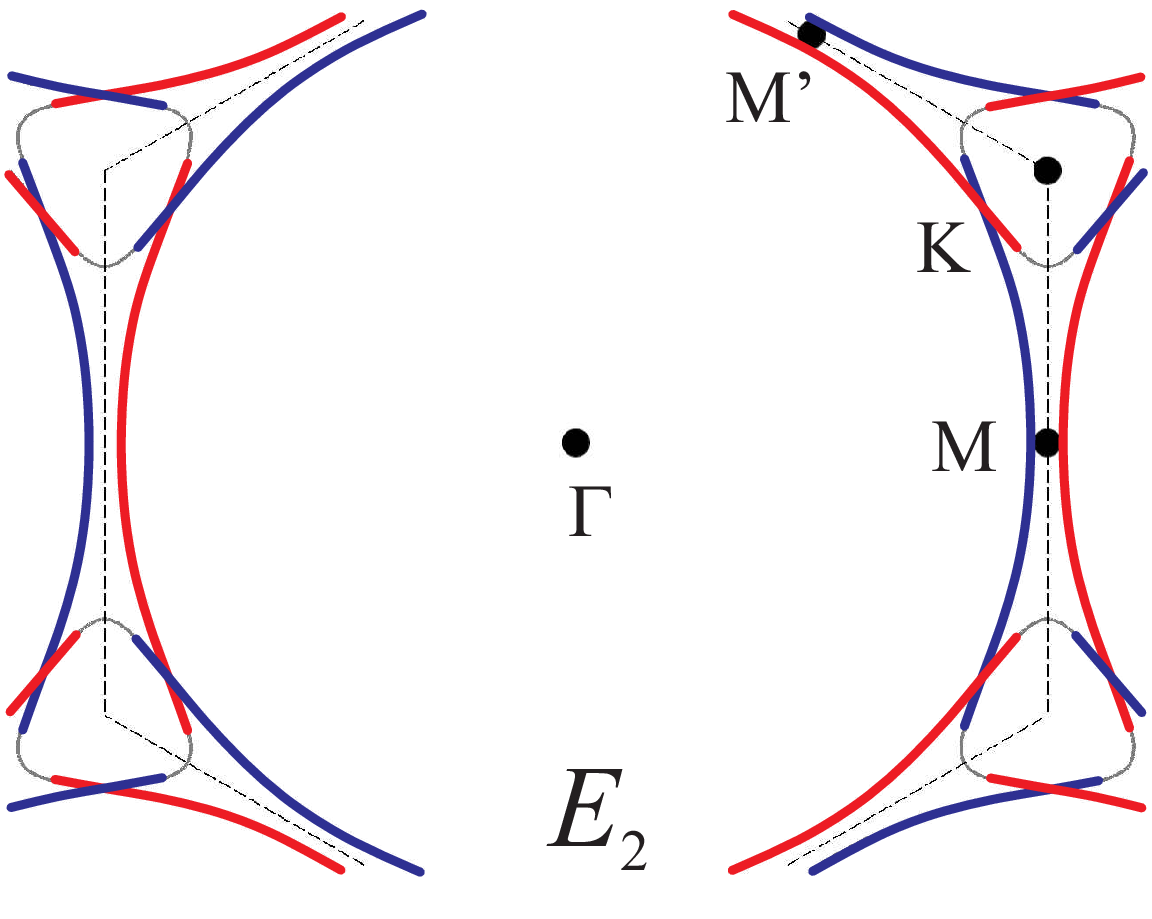}
	\caption{\label{f2} The dispersion of nearly-free electron model Eq.~(\ref{NFE}) along the path shown in the insert. The Fermi surface is plotted for the Fermi level positions at $E_1$ and $E_2$ (the central and right panels). The average spin is indicated by color as in Fig.~\ref{f1}.}
\end{figure*}

The proved theorems and topological arguments can be applied to multidimensional systems. The conductive hexagonal palladium layers in PdCrO$_2$ are described well by a 2D nearly-free-electron model \cite{Kushwaha}. Magnetic spacers CrO$_2$ form an effective field corresponding to a 120$^0$ (three-sublattice) magnetic structure \cite{Takatsu2}. Then, a solution for the Bloch wave functions $\psi$ can be obtained from the following equation \cite{Ashcroft}
\begin{eqnarray}
	\left[ \frac{\hbar^2}{2m}  \big(\mathbf{k} - \mathbf{K}\big)^2 - \mathcal{E} \right] \hat{c}_{\mathbf{k} - \mathbf{K}} +\sum_{\mathbf{K^\prime}} \hat{U}_{\mathbf{\mathbf{K^\prime}} - \mathbf{K}} =0 \label{NFE}
\end{eqnarray} 
where $\mathbf{k}$ is the wave vector within the magnetic Brillouin zone ($\sqrt3 \times \sqrt3$), $\hat{U}_{\mathbf{K}}$ are the Fourier coefficients of effective field, $\hat{c}_{\mathbf{k}}$ and $\hat{U}_{\mathbf{K}}$ have spinor form \cite{Kudasov}. In case of 2D system, the second term in Eq.~(\ref{NFE}) should contain at least two terms to describe properly the band structure in the vicinity of the K points (see Fig.~\ref{f2}). 

Since the model Eq.~(\ref{NFE}) does not contain a spin-orbit coupling, the spin rotation axis $z$ can be chosen arbitrary, and for the sake of simplicity, we direct it along the spacial $z^\prime$ axis, i.e. perpendicular to the plane.       
The Fourier coefficients have a special form in case of a helical 120$^0$ effective field \cite{Kudasov}: $\hat{U}_{\mathbf{K}_{1(2)}}= \gamma h_0 \big( \hat{\sigma}_x \pm i \hat{\sigma}_y \big)$ where $\gamma$ is a complex coefficient. 
It should be mentioned that the spinor Fourier coefficients are abnormal operators \cite{Kudasov}.

The dispersion of the 2D model and the Fermi surface for two positions of the Fermi level are shown in Fig.~\ref{f2}. 
The topology of the dispersion surface is nontrivial. The volume of the extended Brillouin zone is three times larger than that of the magnetic one, therefore the topological index is $n=3$. That is, if we close periodic boundaries of the 2D magnetic Brillouin zone in a torus representation, the dispersion surface wrapping the torus contains three sheets. Intersections of the sheets occur along the Brillouin zone boundaries and  along $\Gamma$-K lines.

Since the model satisfies theorem 2 we again obtain relation (\ref{T1}). In the middle and right panels of Fig.~\ref{f2}, one can see that the opposite arcs of the Fermi surface are predominantly formed by opposite spins. This is an important result because this spin texture suppresses backward nonspin-flip scattering and umklapp electron-phonon scattering  \cite{Kudasov}. The arrows in the middle panel of Fig.~\ref{f2} show transitions between initial and final electron states in the umklapp processes which are suppressed by the spin texture of the Fermi surface. Since at low temperatures an electron-phonon part of resistance in metals with closed Fermi surface is determined by umklapp scattering \cite{ziman}, the resistance occurs strongly suppressed \cite{Kudasov} and the high-conductive state appears. 

In conclusion, the two theorems proved above show that a Kramers-like degeneracy exists in a helical magnetic field. The topology of a band structure in the helical magnetic systems is non-trivial. The traditional topological band theory developed for topological insulators deals with a phase of the wave function and its variation over the Brillouin zone. In the present article, the nontrivial topology stems from the fact that commensurate helical systems have to be simultaneously periodic on the ordinary and generalized translations. This leads to multisheet dispersion of electrons. The specific Kramers-like symmetry  and topology lead to the spin texture of the Fermi surface, which suppresses the backward nonspin-flip scattering and umklapp electron-phonon scattering. As a result, a high-conductivity state appears. This effect is pronounced if a single band crosses the Fermi level (strong topological metal) because otherwise it is masked by interband scattering. This behavior is similar to that of topological surface and edge states in topological insulators \cite{Bansil}. However, in the present work we dealt with bulk states. The magnetic metallic delafossite PdCrO$_2$ is a candidate for a topological metal of this type. It should be mentioned that the band structure in Fig.~\ref{f2} can also reproduce the nonreciprocity of electrons transport under magnetic field observed in this substance \cite{Akaike}. The effect discussed in the present article can be verified by noncollinear calculations within the density functional theory and spin-resolve angle-resolve photoelectron spectroscopy.

\begin{acknowledgments}
%The author thanks A.S. Mel'nikov, A.M. Shikin, and V.N. Petrov for useful discussions. 
The work was supported by National Center for Physics and Mathematics (Project \#7 "Investigations in high and ultrahigh magnetic fields").
\end{acknowledgments}

\bibliography{topmet}

\end{document}